\documentclass{article}

\usepackage{natbib}
\usepackage{rotating}

\usepackage{version}
\usepackage{amsmath} 
\usepackage{url} 
\usepackage{graphicx}
\usepackage{threeparttable}
\newlength{\imgwidth}
\includeversion{onlysubmission}\excludeversion{onlypreprint}
\includeversion{nowordcount}

\title{An explanatory model for food-web structure and evolution}
\author{A. G. Rossberg$^\ast$, H. Matsuda, T. Amemiya, K. Itoh\\
  \normalsize{Yokohama National University, Graduate School of Environment}\\
  \normalsize{and Information Sciences, Yokohama 240-8501, Japan}\\
  \normalsize{$^\ast$Corresponding author. Tel.:
    +81-45-339-4369, fax: +81-45-339-4353}\\
\normalsize{E-mail addresses: rossberg@ynu.ac.jp (A.G.R.),
  matsuda2@ynu.ac.jp  (H.M.),
 }\\\normalsize{amemiyat@ynu.ac.jp (T.A.), itohkimi@ynu.ac.jp (K.I.)}}
\date{}

\begin{document}

\maketitle

\thispagestyle{headings}

\begin{abstract} 
  \setlength{\rightskip}{0pt plus 1fil} 
  Food webs are networks describing who is eating whom in an
  ecological community.  By now it is clear that many aspects of
  food-web structure are reproducible across diverse habitats, yet
  little is known about the driving force behind this structure.
  Evolutionary and population dynamical mechanisms have been
  considered.  We propose a model for the evolutionary dynamics of
  food-web topology and show that it accurately reproduces observed
  food-web characteristic in the steady state.  It is based on the
  observation that most consumers are larger than their resource
  species and the hypothesis that speciation and extinction rates
  decrease with increasing body mass.  Results give strong support to
  the evolutionary hypothesis.
\end{abstract}

\newpage{}


\section{Introduction}
\label{sec:intro}

The complex networks of feeding relations in ecological communities,
food webs, have fascinated researchers for over thirty years
\citep{cohen90:_commun_food_webs_AND_THEREIN}.  Using
carefully collected empirical data%
, previous indications   
\citep{cohen90:_commun_food_webs_AND_THEREIN,havens92:_scale_webs}
could be confirmed that several aspects of the structure of empirical
food webs are reproducible across habitats as diverse as Caribbean
islands, deserts, and lakes
\citep{camacho02:_robus_patter_food_web_struc,
   milo02:_networ_motif,%
   garlaschelli03:_univer,%
  williams00:_simpl,%
  cattin04:_phylog%
}.  Most striking, perhaps, were findings that certain descriptive
model food webs, constructed by following just a few simple rules for
connecting the elements of an abstract species pool, can
quantitatively reproduce large sets of basic statistics of food-web
topology \citep{
  cohen90:_commun_food_webs_AND_THEREIN,%
  williams00:_simpl,%
  cattin04:_phylog%
}.  These models do not yet expose the dynamical processes or
mechanisms by which the structures they describe are formed.  But they
set a standard by which the accuracy of any dynamical explanation
should be measured \citep{williams00:_simpl}.

Two major mechanisms for shaping food webs have been considered.  One
is the selection of those food webs that support stable population
dynamics, because the degree of stability observed in nature is hard
to achieve with simple, randomly assembled food-web models
\citep{
may73:_stabil_complex_eco,yodzis81:stability_real_ecosystems%
}.  The other mechanism is evolutionary dynamics, which is suggested
by the observation that the similarities in the trophic role of
species and their phylogenetic kinship are highly correlated
\citep{cattin04:_phylog}.  Even though there has been plenty of work
on dynamical food-web models incorporating population dynamics,
evolutionary dynamics, or combinations of both \citep[for a review,
see][]{drossel03:_model_food_webs}, non of these models could so far
be shown to reproduce food-web structure to the accuracy of the more
recent descriptive models, namely the \emph{niche model}
\citep{williams00:_simpl} and the \emph{nested hierarchy model}
\citep{cattin04:_phylog}.  Purely evolutionary models
\citep[e.g.,][]{amaral99:_envir_chang_coext_patter_fossil_recor} have
been developed mainly for understanding the dynamics of the total
number of species, and not the detailed food-web structure.  Work on
models combining evolutionary and population dynamics [e.g.
\citet{
caldarelli98:_model_multispec_commun,drossel01:_influen_predat_prey_popul_dynam,yoshida03:_evolut_web_sys%
}] has not yet arrived at conclusive results, one problem being high
model complexity.  A recent systematic tests of a plain population
dynamical model by \citet{montoya03:_topol_webs_real_to_assemb} turned
out clearly negative.

Here we introduce a new model (called \emph{speciation model} below)
for the evolutionary dynamics of food webs, and show that its accuracy
is comparable to the descriptive models.  For an in-depth mathematical
analysis of the model we refer to \citet{rossberg05:spec}.

\section{Model definition}
\label{sec:model}

We model the evolutionary dynamics of an abstract pool of species
belonging to a given habitat, and the topology of the network of
feeding relations between them.
In order to separate the plain macroevolutionary dynamics from the
complex conditions that determine fitness, evolution is modeled as
being undirected.  This simplification is here not meant to be a
statement about nature, but is used as a technique to isolate the
effect of particular processes.
Population dynamics, for example, does not enter the model.

The speciation model is most naturally described in terms of a
continuous-time stochastic process.  (See Appendix~\ref{sec:algorithm}
for a description of our computational implementation.)  Its dynamics
consists of two parts, the dynamics of the community and the dynamics
of the food web connecting the species.  Table~\ref{tab:par-list}
lists the model parameters.

\subsection{Evolution of the species pool}
\label{sec:pool}

Only three kinds of macro-evolutionary processes are taken into
account: speciations and extinctions within a habitat and adaptations
of new species to the habitat.  Anagenesis (evolution without
speciation) is omitted for simplicity.
Since changes in the environment can be fast on evolutionary time
scales, it is for this model more appropriate to conceive a habitat as
a particular set of environmental conditions supporting an ecological
community, rather than just a specific location.

%

Species at higher trophic levels tend to be larger than at lower
levels \citep{memmott00:_predat_size_web,leaper02:_size_in_web}.  Due
to this correlation, a possible correlation between body size and
speciation- and extinction rates leads to correlations between
evolution rates and trophic level.  Arguments in favor of both,
evolution rates increasing \citep{fenchel93:_small_large_spec} and
decreasing \citep{bush93:_santa_rosal} with body size have been put
forward.  Empirically, the question seems to be undecided.  Generally
one would assume that evolution rates are on the average proportional
to some power $M^\alpha$ of the adult body mass $M$.  We note that
$\alpha=0$ is just as difficult to motivate as any other choice.
Based on the comparison of the observed distributions of the number of
links to resources and consumers
\citep{camacho02:_robus_patter_food_web_struc}, it can be argued that
$\alpha<0$ is more plausible in the context of the model proposed here
\citep{rossberg05:spec}.

We associate
each species $i$ with a ``speed parameter'' $s_i$ (triangles in
Fig.~\ref{fig:speciation}) that determines the magnitude
of its evolutionary rates $\sim \exp(s_i)$. 
%
The community dynamics is characterized by the values of the rate
constants for new adaptations $r_1$, speciations $r_+$, and
extinctions $r_-$, the range $[0,R]$ of allowed values for $s$, and
the evolutionary speed/size dispersion $D$.
The complexity of the processes driving evolution is accounted for by
modeling extinctions, speciations, and adaptations of new species to
the habitat as independent random events: For any short time interval
$[t, t+dt]$, the probability that a species $i$ becomes extinct is
$r_- \exp(s_i) \, dt$ and the probability that it speciates is $r_+
\exp(s_i) \, dt$.  When $i$ speciates, a new species $j$ with slightly
different speed parameter $s_j=s_i+\delta$ is introduced into the
community, where $\delta$ is a zero-mean Gaussian random variable with
$\mathop{\mathrm{var}}\delta=D$.  If $s_i+\delta$ exceeds the range
$[0,R]$, $s_j=-(s_i+\delta)$ or $s_j=2 R-(s_i+\delta)$ are used
instead.
The probability that a new species with a speed parameter in a small
range $[s, s+ds]$ adapts to the habitat and joins the community is
$r_1 \exp(s)\, ds\, dt$ for all $s$ between $0$ and $R$.
%

\subsection{Dynamics of the trophic network}
\label{sec:web}

\begin{figure}[h]
  \centering
  \begin{onlypreprint}
    \includegraphics[width=0.7\columnwith,keepaspectratio]{newSpeciation2}
  \end{onlypreprint}
  \begin{onlysubmission}
        \includegraphics[width=\textwidth,keepaspectratio]{newSpeciation2}\\
        \texttt{[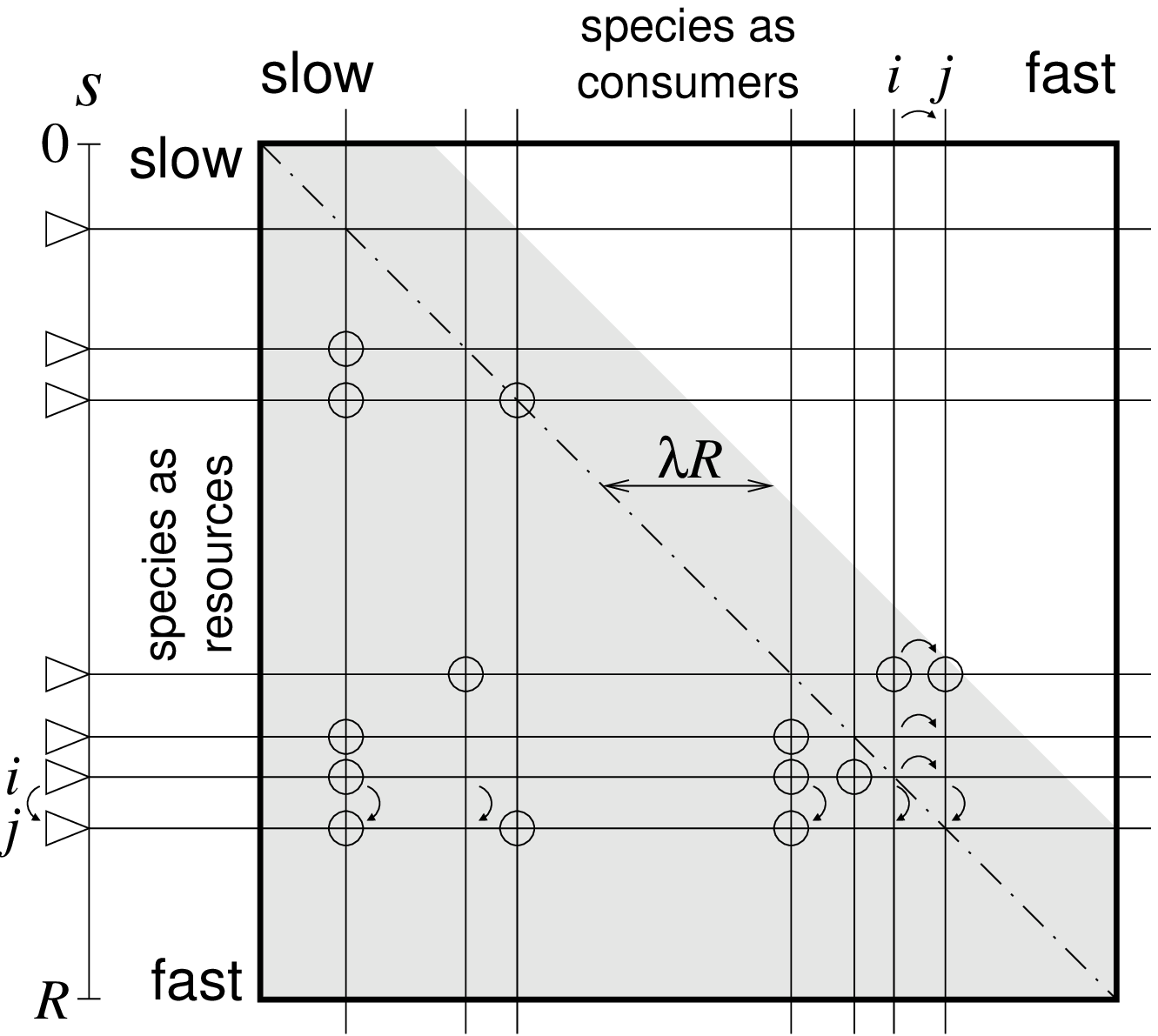]}
  \end{onlysubmission}
  \caption{\baselineskip24pt 
    Illustration of the speciation model\newline  The food web  is
    represented by a connection matrix (large square).  Horizontal
    lines represent species as resources, vertical lines the same set of
    species as consumers.  Trophic links are indicated by circles at
    the intersection points.  The $s$-axis marks the evolution rate of
    each species.  The process shown is a speciation. Species $j$
    evolves from species $i$.  Most trophic links of $j$ are copied
    from $i$, but some are modified.}
\label{fig:speciation}
\end{figure}

The food web is described by the topology of the network of directed
trophic links connecting species.  Parameters determining its dynamics
are the loopiness $\lambda$, the raw connectivity $C_0$, and the
reconnection probability $\beta$.
Following \citet{cohen90:_commun_food_webs_AND_THEREIN},
\citet{williams00:_simpl}, and \citet{cattin04:_phylog}, we take the
fact into account that consumers are typically larger than their
resource species.  Empirical data show that they are, parasites
excluded, at least not much smaller in terms of adult body mass
\citep{memmott00:_predat_size_web,leaper02:_size_in_web}.  We assume
that there is an upper limit $h$ for the ratio of the adult body sizes
of resource and consumer species.  Resource species are at most $h$
times larger than their consumers.  Making use of the assumption that
evolution rates $\sim\exp(s)$ scale as $M^{\alpha}$ with $\alpha<0$,
the difference of the speed parameter $s$ of a resource and its
consumer has an upper limit approximately given by $\Lambda:= \left|
  \alpha \right|\times\ln h$. The degree by which the trophic ordering
is violated within the species pool is expressed by the loopiness
parameter $\lambda=\Lambda/R$.  Summarizing, size constraints restrict
the set of \emph{possible consumers} of a species $i$ to the species
$l$ with $s_l<s_i+\lambda\,R$ and the \emph{possible resources} to the
species $h$ with $s_h>s_i-\lambda\,R$ where $0\le\lambda\le1$ (shaded
area in Fig.~\ref{fig:speciation}).  Thus, cannibalism and trophic
loops \citep{polis91:_compl_des_web} can occur.

When a new species $k$ is adapting to the habitat, a possible consumer
of $k$ becomes its consumer with probability $C_0$ and, with the same
probability, a possible resource of $k$ becomes its resource.  When a
species $i$ speciates (Fig.~\ref{fig:speciation}), the connections of
the descendant species $j$ are determined in three steps.  First, all
possible resources and consumers of $i$ that are also possible
resources and consumers of $j$ are linked to $j$.  Possible resources
and consumers of $j$ that are not possible consumers or resources of
$i$ are linked with probability $C_0$.  Then, for each possible
consumer or resource of $j$, the information whether it is connected
is discarded with probability $\beta$.  Finally, a link is established
with probability $C_0$ to any consumer or resource for which the
information was discarded.  Species that are both possible resources
and possible consumers of $j$ are treated like two species: a consumer
and a resource.  This procedure ensures that the average ratio of
possible links to realized links remains at $C_0$, in line with the
premise that evolution is essentially undirected.
The steady-state properties of the model are independent of the
initial conditions.
Approximations for some basic properties can be obtained analytically
\citep{rossberg05:spec}.  For example, the expected density of species
has a constant value $r_1/(r_--r_+)$ along the $s$ axis.  Thus, the
observed broad distributions of logarithmic body masses
\citep[e.g.][]{fenchel93:_small_large_spec} are here modeled by a
simple boxcar function.  The number of species $S$ is on the average
$\overline{S}=R r_1/(r_--r_+)$ and, with $L$ denoting the number or
links, the average directed food-web connectivity defined as $C=L/S^2$
is approximately $\overline{C} \approx C_0\, (1+2
\lambda-\lambda^2)/2$.  The fraction of species that entered the
habitat by speciation is $r_+/r_-$, the rest ($1-r_+/r_-$) entered as
new adaptations.

\section{Model verification}
\label{sec:empirical}

The speciation model was compared with seven of the best available
empirical webs (see Table~\ref{tab:parameters})%
, all of which exclude parasites, in line with our model.
Collecting food-web data is not at all an easy task, and the data is
sometimes criticised for being incomplete or inconsistent.  But it was
for this kind of data that the regularities mentioned in
Section~\ref{sec:intro} have been found, which suggests that the data
does carries substantial information characteristic for food webs.

\begin{figure}[h]
  \centering
  \begin{onlypreprint}
    \includegraphics[width=0.9\textwidth,keepaspectratio]{samples}
  \end{onlypreprint}
  \begin{onlysubmission}
    \includegraphics[width=0.8\textwidth,keepaspectratio]{samples}\\
        \texttt{[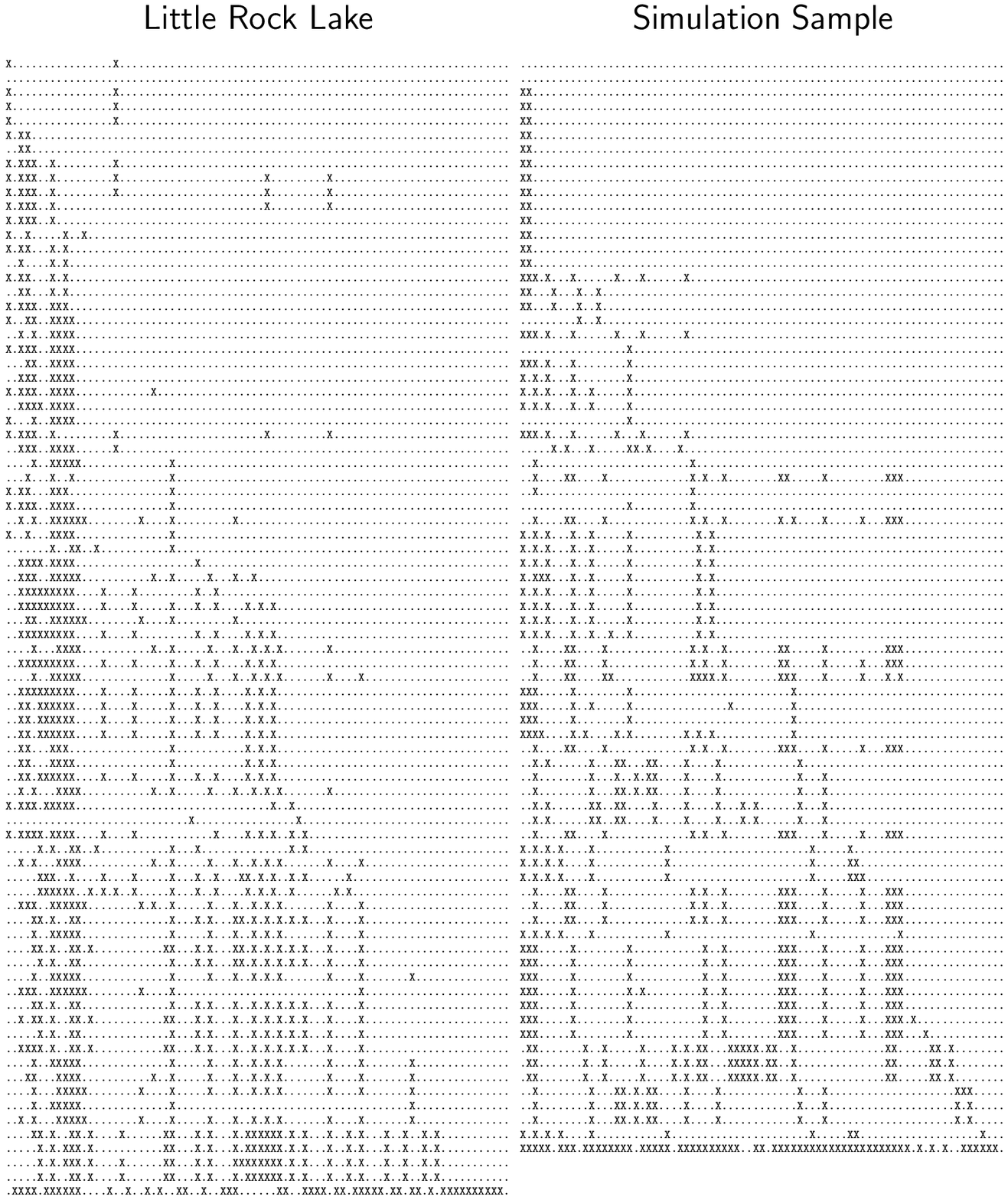]}
  \end{onlysubmission}
  \caption{Examples for connection matrices of food webs\newline
    The connection matrix of the food webs of Little Rock Lake
    \citep{martinez91:_artif_attr} and a simulation sample, both after
    species lumping (see Appendix~\ref{sec:properties} for lumping
    procedure).  Model parameters are as in
    Table~\ref{tab:parameters}, Little Rock.  Each 'X' indicates that
    the species corresponding to the column eats the species
    corresponding to the row.  The ordering of the species is such as
    to minimize the number of upper diagonal links and otherwise
    random.}
  \label{fig:matrix}
\end{figure}

A simple, visual comparison of the connection matrix of the food web
of Little Rock Lake \citep{martinez91:_artif_attr} with a model sample
(Fig.~\ref{fig:matrix}) shows that the model reproduces the
characteristic mixture of randomly scattered blocks and isolated links
found in empirical data.  The empirical web has several omnivorous
species of high trophic level, which leads to a high density of links
near the left edge of the connection matrix.  Such isolated, dense
blocks are only occasionally found in simulations.  They are not
typical for empirical webs either.

A systematic model verification was done based on twelve commonly used
food-web statistics (see Appendix~\ref{sec:properties}).  We kept the
parameters $R=\ln 10^4$ and $D=0.0025$ fixed, since their precise
value has little effect ({doubling either $D$ or $R$ changes the
  $\chi^2$ values computed below by less than one.}), and set $r_-=1$
without loss of generality.  Depending on $r_+$, $\lambda$, and
$\beta$ (see below), parameters $r_1$ and $C_0$ were chosen so that
estimates of $\overline{S}$ and $\overline{C}$ after data
standardization (Appendix~\ref{sec:properties}) match with the species
abundance $S_\text{e}$ and connectivity $C_\text{e}$ of the empirical
webs.  Maximum likelihood estimates (see Appendix~\ref{sec:stats})
assuming a multivariate Gaussian distribution for the twelve food-web
statistics were used for the remaining model parameters $r_+$,
$\lambda$, and $\beta$ (Table~\ref{tab:parameters}).  As a benchmark,
food-web statistics were also computed for the niche model
\citep{williams00:_simpl} and for the nested hierarchy model
\citep{cattin04:_phylog}.  Neither has free parameters.  As is shown
in Fig.~\ref{fig:properties}, there is good overall agreement between
empirical and model data for all models.


As a systematic measure for the goodness of fit, we computed the
$\chi^2$ statistics given by
\begin{align}
  \label{chi2}
  \chi^2=(\mathbf{v}_\text{e}-\overline{\mathbf{v}})^T \mathbf{C}^{-1}
  (\mathbf{v}_\text{e}-\overline{\mathbf{v}}),
\end{align}
where $\mathbf{v}_\text{e}$ denotes the vector of the twelve
food-web statistics for an empirical web, and $\overline{\mathbf{v}}$
and $\mathbf{C}$ are the corresponding model mean and covariance
matrix (Appendix~\ref{sec:stats}).
In terms of $\chi^2$ statistics (Table~\ref{tab:parameters}) the
speciation model ($\chi^2_\text{S}$) is more accurate than the niche
($\chi^2_\text{N}$) and the nested hierarchy model
($\chi^2_\text{H}$), even when differences in the number of
statistical degrees of freedom (DOF $=$ number components of
$\mathbf{v}$ minus number of fitting parameters) are taken into
account. For example, the overall $p$ value for either niche or nested
hierarchy model is less than $10^{-24}$ times that of the speciation
model.  In all but one case, the speciation model yields the lowest
$\chi^2$.  This is partially due to inherently larger model variances
and partially due to a better accuracy of the model averages: in 46
out of 84 cases the average of the speciation model is closest to the
empirical value ($p<0.001$).  Smaller webs are modeled particularly
well.  For the five food webs with $S_e$ up to 35 (after species
lumping) all $\chi^2_\text{S}$ values lie within the $5\%$ confidence
interval $\chi^2_\text{S}<16.9$; and so does their sum $\sum
\chi^2_\text{S}=50.5 < 61.7$.

\section{Discussion}

In order to exclude the fallacies that the surprisingly good
quantitative fit of the speciation model (1) does not actually depend
on evolutionary processes or (2) is merely a consequence of the data
standardization and evaluation procedures deployed, the model was also
evaluated with $r_+=0$ fixed, i.e.\ with all speciation processes
inhibited.  In this case the parameter $\beta$ also does not have any
effect, so that the only remaining fitting parameter is $\lambda$.
The model fits worsened considerably.  We obtained $\chi^2$ values
ranging from $65$ (Chesapeake Bay) to over 1000 (Ythian Estuary,
Little Rock Lake).

Of course this does not prove that there is no other explanation for
the observations, but it shows that, for the good fit of the model,
the speciation events are crucial.  Two effects of the evolutionary
dynamics are probably particularly important: (1) It leads to
approximately exponential distributions in the number of resources of
a species \citep{rossberg05:spec}, distributions that are similar to
those obtained with the niche model by prescribing a beta distribution
for the ``niche width''
\cite{camacho02:_analytic_food_webs,camacho02:_robus_patter_food_web_struc}.
(2) Related species have similar sets of consumers and resources, an
effect that the algorithm of the nested hierarchy model
\citep{cattin04:_phylog} was set up to mimic.

The speciation model is at least as accurate in its
predictions as the niche model, which was itself a big improvement
over the, by now historical, cascade model
\citep{cohen90:_commun_food_webs_AND_THEREIN}.  The speciation model is
the first that involves a mechanistic explanation, confirming the
earlier hypothesis \citep{cattin04:_phylog} that the structure of food
webs can be understood as the effect of a sequence of evolutionary
events.
Recent theoretical and empirical results
\citep{mccann00:_diversity_stability_review} show that the population
dynamics in complex food webs can be stabilized by mechanisms such as
weak links and adaptive foraging
\citep{kondoh03:_forag_adapt_relat_between_food}.  This indicates that
population-dynamical restrictions on food-web structure are less
severe than one might first assume, making room for evolutionary
dynamics to leave its traces.

\section{Acknowledgements}
\label{sec:thanks}

The authors express their gratitude to N.~D.~Martinez and
his group for making their food-web database available and The 21st
Century COE Program ``Bio-Eco Environmental Risk Management'' of the
Ministry of Education, Culture, Sports, Science and Technology of
Japan for generous support.

\clearpage{}

\newpage

\begin{nowordcount}

\appendix

\makeatletter{}
\@addtoreset{equation}{section} 
\makeatother{}
\renewcommand{\theequation}{\thesection.\arabic{equation}}

\section*{Appendices}
\label{appendices}

\section{An Efficient Implementation of the Speciation Model}
\label{sec:algorithm}

\noindent An evaluation of the speciation model based on
a straightforward discretization of speed parameters $s$ and time
$t$ would require excessive amounts of computation time.  Here the
data structures and algorithms that were used in our, more efficient,
implementation are described.

\subsection{From Poisson processes to event sequences}

The main loop of our algorithm advances the dynamics of the food webs
by one evolutionary event at each iteration.  The question which event
comes next is decided based on the following considerations:
Call the quantity $\sigma_i:=\exp(s_i)$ the \emph{raw evolution rate},
or simply \emph{evolution rate} of species $i$.  We first note that
the probability for new species with an evolution rate in the range
$[\sigma,\sigma+d\sigma]$ to adapt is
\begin{align}
  \label{sigma}
  r_1\,\exp(s)\,ds\,dt=r_1\,\sigma\, \frac{ds}{d\sigma}\,
  d\sigma\,dt=r_1 d\sigma\,dt.
\end{align}
Thus, the total rate at which species adapt to the habitat is simply
\begin{align}
  \label{R1}
  \nu_1:=\int_{\sigma_\text{min}}^{\sigma_\text{max}} r_1 d\sigma =
  r_1\,({\sigma_\text{max}}-{\sigma_\text{min}})
\end{align}
with $\sigma_\text{min}:=\exp(0)=1$ and $\sigma_\text{max}:=\exp(R)$.
Since all possible events are assumed to be statistically independent,
the probability that \emph{any} new-adaptation, extinction, or
speciation event occurs in the infinitesimal time interval $[t,t+dt]$
is $\nu_\text{tot} dt$ with
\begin{align}
  \label{Rtot}
  \nu_\text{tot}:= \nu_1 + \left(r_- +
    r_+\vphantom{\sum}\right)\,\sum_{i} \sigma_i,
\end{align}
where the sum runs over the indices of all species currently in the
species pool.  The time $\tau$ to the next event follows the
distribution
\begin{align}
  \label{between-time}
  P(\tau)=\nu_\text{tot}\,\exp(-\nu_\text{tot} \tau).
\end{align}

If the probability for some event $X$
to occur in an interval $[t,t+dt]$ is $\nu_X\,dt$, then the
probability for event $X$ to occur as the next event is
\begin{align}
  \label{pnext}
  \frac{\nu_X}{\nu_\text{tot}}.
\end{align}
To see this, consider first the situation with only two kinds of
events: event $A$ occurring at rate $\nu_A$ and event $B$ at rate
$\nu_B$ (both are Poisson processes).  The joint probability
distribution for the time $\tau_A$ to the next event $A$ and the time
$\tau_B$ to the next event $B$ is
\begin{align}
  \label{Pab}
  P_{AB}(\tau_A,\tau_B)=\nu_A\,\exp(-\nu_A\tau_A)\times
  \nu_B\,\exp(-\nu_B\tau_B).
\end{align}
The probability that $A$ occurs before $B$ is
\begin{align}
  P \left[ \tau_A < \tau_B \right]=&\int_0^\infty
  d\tau_A\int_{\tau_A}^\infty d\tau_B
  P_{AB}(\tau_A,\tau_B)\notag\\
  \label{AbeforeB}
  =&\int_0^\infty d\tau_A \nu_A \exp(-\nu_A\tau_A-\nu_B\tau_A)\\
  =&\frac{\nu_A}{\nu_A+\nu_B}.\notag
\end{align}
The probability (\ref{pnext}) is obtained by letting event $A$ be
event $X$ and event $B$ be any other event.

For the sake of formal simplicity the range of allowed values for the
speed parameter $s$ is reduced from $[0,R]$ to the half open interval
$[0,R[$ without affecting numerical results.  In order to decide which
event comes next in the sequence of events, a random number $\rho$
equally distributed in the range $[0,\nu_\text{tot}[$ is generated.
If $\rho<\nu_1$, the next event is  taken to be a new adaptation and
in view of Eq.~(\ref{sigma}) the evolution rate of the new species is
chosen as a random number equally distributed in the range
$[\sigma_\text{min},\sigma_\text{max}[$.  If $ \nu_1\le\rho$, the next
event is taken to be either the extinction or a speciation of the
species $i$ that satisfies
\begin{align}
  \label{chose-i}
  \left(r_- +
    r_+\vphantom{\sum}\right)\,\sum_{\sigma_j>\sigma_i}\sigma_j \le
  \rho -\nu_1 < \sigma_i+\left(r_- +
    r_+\vphantom{\sum}\right)\,\sum_{\sigma_j>\sigma_i}\sigma_j.
\end{align}
%
Another random number is used to decide if the species $i$ goes
extinct [probability $r_-/(r_-+r_+)$] or speciates.

\subsection{Data structures for keeping track of the network}

The connectivity between species is stored in a matrix of bits
$\{b_{ij}\}$ with $b_{ij}=1$ is $j$ eats $i$ and $b_{ij}=0$ otherwise.
In order to avoid having to re-arrange the full matrix $b_{ij}$ with
every event, the number of rows and columns of $\{b_{ij}\}$ is chose
much larger than the actual number of species is the pool.
Indices are dynamically allocated with every new adaptation or
speciation, and deallocated upon extinction.  Unused indices are kept
in a stack data-structure.
The determination of the connectivity follows directly the model
definition.



\subsection{Sampling the steady state}

It is important to notice that the main loop of the algorithm, which
simulates one evolutionary event per iteration, does not proceed with
constant speed along the time axis $t$.  Episodes with many species
require more iterations per unit time than episodes with few species.
In order to sample the steady state evenly in $t$, the an estimate $\hat t$
of $t$ has to be used, since the true time between events is left
undetermined with the method described above.  Such an estimate can be
obtained by starting with $\hat t=0$ and incrementing $\hat t$ by the
current expectation value of the time between events
$\tau_\text{tot}:=\nu_\text{tot}^{-1}$ at each iteration.

The portion of the food web that involves small species with large
$\sigma$ reaches the steady state much faster than the portion that
involves only large, slow species (small $\sigma$).  In the initial
stage of the simulation, it is therefore sufficient to account for the
portion involving only slow species.  In practice, this is achieved by
setting the effective value $R'$ of the parameter $R$ to a low value
$R_\text{start}$ (e.g., $R_\text{start}=1$) at the beginning of the
simulation and increasing $R'$ after each iteration by a small amount
proportional to $\tau_\text{tot}$.  When $R'$ has reached $R$, the
simulation is continued for some time, and then a steady-state
food web is sampled.  For the next steady-state sample, the complete
food web is discarded, and the simulation is restarted with a
$R'=R_\text{start}$.  We started each run with an empty species pool.

In order to check if the relaxation time is long enough to reach the
steady state, we verified that, for the steady-state samples taken,
the density of species along the $s$ axis has the theoretically
expected constant value $r_1/(r_--r_+)$ for $D=0$.

\section{Food-web properties}
\label{sec:properties}

The food-webs properties that were used to characterize and compare
empirical and model webs were: the clustering coefficient
\citep{camacho02:_robus_patter_food_web_struc,
  dorogovtsev02:_evolut_networ} (\textit{Clust} in
Figure~\ref{fig:properties}); the fractions of cannibalistic species
\citep{williams00:_simpl} (\textit{Cannib}) and species without
consumers \citep{cohen90:_commun_food_webs_AND_THEREIN} (\textit{T},
top predators); the relative standard deviation in the number of
resource species \citep{schoener89:_food_webs} (\textit{GenSD},
generality s.d.) and consumers \citep{schoener89:_food_webs}
(\textit{VulSD}, vulnerability s.d.); the web average of the maximum
of a species' Jaccard similarity \citep{jaccard08:_nouvel_florale}
with any other species \citep{williams00:_simpl} (\textit{MxSim}); the
fraction of triples of species with two or more resources, which have
sets of resources that cannot be ordered to be all contiguous on a
line \citep{cattin04:_phylog} (\textit{Ddiet}); the average
\citep{cohen90:_commun_food_webs_AND_THEREIN} (\textit{aChnLg}),
standard deviation \citep{martinez91:_artif_attr} (\textit{aChnSD}),
and average per-species standard deviation
\citep{goldwasser93:_const_carib_web} (\textit{aOmniv}, omnivory) of
the length of food chains, as well as the $\log_{10}$ of their total
number \citep{martinez91:_artif_attr} (\textit{aChnNo}), with the prefix
\textit{a} indicating that these quantities were computed using the
fast, ``deterministic'' Berger-Shor approximation
\citep{berger90:_approx_subgraph} of the maximum acyclic subgraph
(MAS) of the food web. The number of non-cannibal trophic links not
included in the MAS was measured as \textit{aLoop}. When the output
MAS of the Berger-Shor algorithm was not uniquely defined, the average
over all possible outputs was used. 

The statistics were computed after standardization of food webs to
allow a consistent comparison of data: Since in many records of
empirical food webs the lowest trophic level is poorly resolved, all
species without resource species were lumped to a single ``trophic
species'' after dropping disconnected species and before the usual
lumping of trophically equivalent species
\citep{cohen90:_commun_food_webs_AND_THEREIN}, for both empirical and
model data.

\section{Maximum Likelihood Estimates, 
  Goodness of Fit, and Confidence Intervals}
\label{sec:stats}

Maximum likelihood estimation of model parameters and a systematic
characterization of the goodness of fit ideally require knowledge of
the joint probability distribution of the food-web statistics in the
steady state for any given set of model parameters.  We used the
methods described below to obtain satisfactory approximation in a
computationally feasible way.

Since species number $S$ and directed connectivity $C$ tend to
fluctuate strongly in the steady state of the speciation model, it is
impracticable to restrict sampling to those webs with $S$ and $C$ in
close vicinity of the corresponding empirical values $S_\text{e}$ and
$C_\text{e}$ as done by \citet{williams00:_simpl,cattin04:_phylog}.
Instead, all webs with values of $S$ and $C$ differing by less than
$30\%$ from $S_\text{e}$ and $C_\text{e}$ were used for the
statistical analysis.  In order to estimate the steady-state mean
$\overline{\mathbf{v}}$ and covariance matrix $\mathbf{C}$ of the
twelve food-web statistics \emph{conditional to} $S=S_\text{e}$,
$C=C_\text{e}$, we first computed steady-state mean
$\overline{\mathbf{u}}$ and covariance matrix $\mathbf{D}$ of the
vector
\begin{align}
  \mathbf{u}=
  \left(
    \begin{matrix}
      \mathbf{v}\\
      \mathbf{w}
    \end{matrix}
  \right),
\end{align}
consisting of the twelve food web statistics $\mathbf{v}$ and the
vector $\mathbf{w}=(S,C)^T$ from $N=1000$ samples within the
$\pm30\%$~range.
Using $\mathbf{w}_\text{e}=(S_\text{e},C_\text{e})^T$ and components
of $\overline{\mathbf{u}}$ and $\mathbf{D}^{-1}$ written in the block
matrix forms
\begin{align}
  \label{cp}
  \overline{\mathbf{u}}=
  \left(
    \begin{matrix}
      \overline{\mathbf{u}}_\mathbf{v}\\
      \overline{\mathbf{u}}_\mathbf{w}
    \end{matrix}
  \right),\quad
  \mathbf{D}^{-1}=
  \left(
    \begin{matrix}
      \mathbf{M}_{\mathbf{v}\mathbf{v}} & 
      \mathbf{M}_{\mathbf{v}\mathbf{w}} \\
      \mathbf{M}_{\mathbf{w}\mathbf{v}} & 
      \mathbf{M}_{\mathbf{w}\mathbf{w}} & 
    \end{matrix}
  \right),
\end{align}
estimates for
$\overline{\mathbf{v}}$ and $\mathbf{C}$ can be obtained as
\begin{subequations}
  \label{projection}
\begin{align}
  \overline{\mathbf{v}}=&\overline{\mathbf{u}}_\mathbf{v}+
  \mathbf{M}_{\mathbf{v}\mathbf{v}}^{-1}\,\mathbf{M}_{\mathbf{v}\mathbf{w}}\,
  \left( \overline{\mathbf{u}}_\mathbf{w}-\mathbf{w_\text{e}}
  \right),\\
  \mathbf{C}=&\mathbf{M}_{\mathbf{v}\mathbf{v}}^{-1}.
\end{align}
\end{subequations}
The linear projection~(\ref{projection}) is straightforwardly verified
to be exact when $\mathbf{u}$ and, as a result, $\mathbf{v}$ have
multivariate Gaussian distributions and \hbox{$N\to\infty$}.  Maximizing the
corresponding likelihood function for the empirical statistics
conditional to $S=S_\text{e}$, $C=C_\text{e}$,
\begin{align}
  \label{lh}
  \frac{1}{\sqrt{(2\pi)^{12}
      \left|
        \mathbf{C}
      \right|}}\exp\left[-\frac{1}{2}
  (\mathbf{v}_\text{e}-\overline{\mathbf{v}})^T \mathbf{C}^{-1}
  (\mathbf{v}_\text{e}-\overline{\mathbf{v}})\right],
\end{align}
is equivalent to minimizing
\begin{align}
  \label{chi-det}
  \chi^2+\mathop{\mathrm{ln}}
  \left|
    \mathbf{C}
  \right|
\end{align}
with $\chi^2$ as given by Eq.~(\ref{chi2}) in the main text.  The
maximum likelihood estimates for the model parameters $r_+$,
$\lambda$, and $\beta$ were obtained by minimizing~(\ref{chi-det})
using the global optimization algorithm of
\citet{schonlau98:_global_versus_local_searc_const}.


When the number of samples $N$ used for estimating
$\overline{\mathbf{v}}$ and $\mathbf{C}$ in the formula for $\chi^2$
(main text) is small, statistical errors of these estimates broaden
the probability distribution of $\chi^2$ beyond the usual
$\chi^2$-distribution.  In the simplest case $\chi^2$ (precisely
$N\chi^2$) then follows Hotelling's $T^2$ distribution, for which
analytic expressions are known.  We decided to use the infinite
sample-size confidence intervals (in $95\%$ of all cases $\chi^2<16.9$
for $9$ DOF and $<61.7$ for $9\times 5$ DOF) instead of the finite
sample size intervals for $N=1000$ ($<17.1$ and $<65.1$).
This is the more conservative choice and, since our procedure for
estimating $\overline{\mathbf{v}}$ and $\mathbf{C}$ is different from
the one assumed for Hotelling's $T^2$ distribution, it is not clear if
these finite-sample-size corrections would still apply.

\bibliographystyle{elsart-harv} 
\bibliography{/home/axel/bib/bibview}

\end{nowordcount}

\begin{onlysubmission}
  \newpage{}
\end{onlysubmission}
\begin{onlypreprint}
  \thispagestyle{empty} \pagestyle{empty}
\end{onlypreprint}


\newpage 

\begin{table*}[h] 
  \begin{threeparttable}
    \caption{Empirical food webs, fitting parameters, and goodness of fit}
  \begin{tabular}{llrrrrrrrrrr}
    Food Web & key &\multicolumn{1}{c}{ $S_\text{e}$\tnote{a}} & \multicolumn{1}{c}{$C_\text{e}\tnote{a}$} & \multicolumn{1}{c}{$\displaystyle\frac{r_+}{r_-}$} & \multicolumn{1}{c}{$r_1$} & \multicolumn{1}{c}{$\lambda$ }&\multicolumn{1}{c}{ $C_0$} &
    \multicolumn{1}{c}{$\beta$} & \multicolumn{1}{c}{ $\chi^2_\text{N}$\tnote{b}} & \multicolumn{1}{c}{  $\chi^2_\text{S}\tnote{c}$} &  \multicolumn{1}{c}{ $\chi^2_\text{H}\tnote{d}$} \\
    \hline
    \multicolumn{10}{r}{{
        number of statistical DOF:} 11\tnote{e}} & 9 & 12 \\
    \hline
    Bridge Brook Lake & BB& 15 & 0.28 & 0.91 & 0.17 & 0.12\phantom{5} & 0.37 & 0.059 & 43 & 13.9 & 44 \\
\multicolumn{12}{l}{\citep{havens92:_scale_webs}}\\
    Skipwith Pond &Sk& 25 & 0.32 & 0.93 & 0.21 & 0.009 & 0.53 & 0.012 & 83 & 11.4 & 64 \\
\multicolumn{12}{l}{\citep{warren89:_spatial_freshw_web}}\\
    Coachella Desert &Co& 27 & 0.34 & 0.96 & 0.13 & 0.006 & 0.58 & 0.014 & 37 & 3.9 & 152 \\
\multicolumn{12}{l}{\citep{polis91:_compl_des_web}}\\
    Chesapeake Bay &Ch & 27 & 0.08 & 0.96 & 0.21 & 0.25\phantom{3} & 0.06 & 0.029 & 14 & 7.5 & 23 \\
\multicolumn{12}{l}{\citep{baird89:_chesap_bay}}\\
    St.~Martin Island & SM& 35 & 0.14 & 0.80 & 0.92 & 0.000 & 0.23 & 0.034 & 17 & 13.8 & 12 \\
\multicolumn{12}{l}{\citep{goldwasser93:_const_carib_web}}\\
    Ythan Estuary &Yth& 78 & 0.06 & 0.95 & 0.67 & 0.001 & 0.08 & 0.040 & 58 & 36.8 & 82 \\
\multicolumn{12}{l}{\citep{hall91:_food_rich_web}}\\
    Little Rock Lake & LR& 80 & 0.15 & 0.99 & 0.13 & 0.025 & 0.16 & 0.006 & 52 & 24.2 & 165 \\
\multicolumn{12}{l}{\citep{martinez91:_artif_attr}}\\
  \end{tabular}
  \begin{tablenotes}
  \item [a] after species lumping (Appendix~\ref{sec:properties})
  \item [b] niche model \citep{williams00:_simpl}
  \item [c] speciation model (this work)
  \item [d] nested hierarchy model \citep{cattin04:_phylog}
  \item [e] the statistic \textit{Ddiet}
    (Appendix~\ref{sec:properties}), which is always zero for the
    niche model, was excluded
  \end{tablenotes}
\end{threeparttable}
\label{tab:parameters}
\end{table*}

\begin{table}[h]
\caption{Model parameters and their theoretical range}
\begin{tabular}{lcc}
  model parameter & & range\\
  \hline
  rate constant for new adaptations & $r_1$ & $r_1>0$ \\
  rate constant for speciations& $r_+$ & $r_+\ge 0$\\
  rate constant for extinctions& $r_-$ & $r_- > r_+$\\
  total range of evolution rates & $R$ & $R \ge 0$\\
  rate/size dispersion constant & $D$ & $D\ge0$ \\
  loopiness & $\lambda$ & $0\le \lambda \le 1$ \\
  raw connectivity  & $C_0$ &  $0\le C_0  \le 1$ \\
  re-connection probability & $\beta$ & $0\le \beta \le 1$
\end{tabular}\centering
\label{tab:par-list}
\end{table}

\clearpage{}

\newpage{}

\begin{figure*}[h]
  \centering
  \begin{onlypreprint}
    \includegraphics[width=\textwidth,keepaspectratio]{graph}
  \end{onlypreprint}
  \begin{onlysubmission}
    \includegraphics[width=\textwidth,keepaspectratio]{graph}\\
        \texttt{[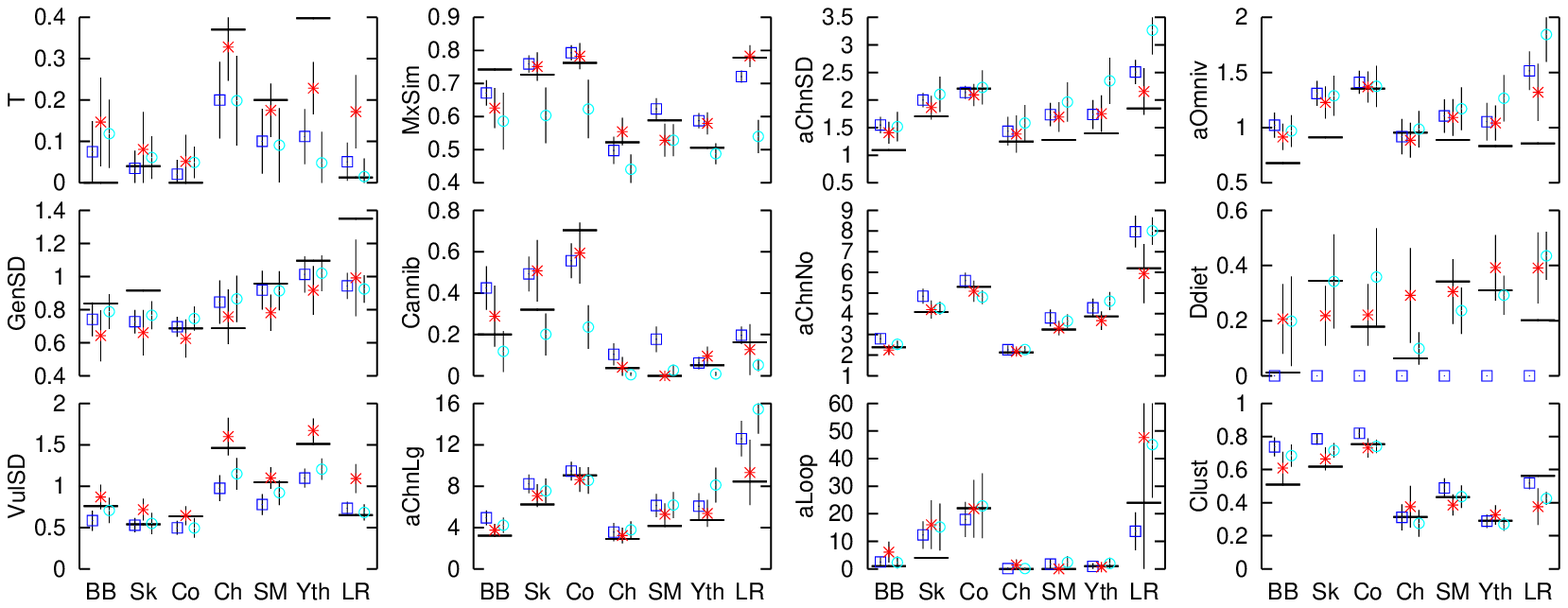]}
  \end{onlysubmission}
    \caption{\baselineskip24pt
      Comparison of model and empirical data\newline The graphs
      display the values of twelve food-web statistics
      (Appendix~\ref{sec:properties}) of seven food webs (key in
      Table~\ref{tab:parameters}) obtained from the speciation model
      (red stars), in comparison with the empirical data (horizontal
      lines), the niche model \citep{williams00:_simpl} (blue
      squares), and the nested hierarchy model
      \citep{cattin04:_phylog} (blue circles).  Vertical lines
      indicate model standard deviations.}\ 
    \label{fig:properties}
\end{figure*}

\clearpage

 \end{document}